\documentclass[twocolumn,superscriptaddress,floatfix,preprintnumbers, nofootinbib,hyperref]{revtex4} %
\usepackage{graphicx}
\usepackage{rotating}
\usepackage{epsfig}
\usepackage{bm}
\usepackage[usenames,dvipsnames]{xcolor}

\usepackage{color}

\def\he4{$^4$He}
\def\h2{$^2$H}

\newcommand{\lesssim}{\,\rlap{\lower3.7pt\hbox{$\mathchar\sim$}}
\raise1pt\hbox{$<$}\,}

\begin{document}

\title{Constraints on the coupling with photons of heavy axion-like-particles \\ from Globular Clusters
}

\author{Pierluca Carenza}
\affiliation{Dipartimento Interateneo di Fisica ``Michelangelo Merlin'', Via Amendola 173, 70126 Bari, Italy}
\affiliation{Istituto Nazionale di Fisica Nucleare - Sezione di Bari, Via Orabona 4, 70126 Bari, Italy}

\author{Oscar Straniero}
\affiliation{INAF, Osservatorio Astronomico d'Abruzzo, 64100 Teramo, Italy}

\author{Babette~D\"obrich}
\affiliation{CERN, Esplanade des Particules 1, 1211 Geneva 23, Switzerland}

\author{Maurizio~Giannotti}
\affiliation{Physical Sciences, Barry University, 11300 NE 2nd Ave., Miami Shores, FL 33161, USA}

\author{Giuseppe Lucente }
\affiliation{Dipartimento Interateneo di Fisica ``Michelangelo Merlin'', Via Amendola 173, 70126 Bari, Italy}

\author{Alessandro~Mirizzi} 
\affiliation{Dipartimento Interateneo di Fisica ``Michelangelo Merlin'', Via Amendola 173, 70126 Bari, Italy}
\affiliation{Istituto Nazionale di Fisica Nucleare - Sezione di Bari, Via Orabona 4, 70126 Bari, Italy}

\date{\today}

\begin{abstract}

We update the globular cluster bound on massive ($m_a$ up to a few 100 keV) axion-like particles (ALP) interacting with photons. 
The production of such particles in the stellar core is dominated by the Primakoff $\gamma + Ze\to Ze +a$ and by
the photon coalescence process $\gamma+\gamma\to a$.
The latter, which is predominant at high masses, was not included in previous estimations.
Furthermore, we account for the possibility that axions decay inside the stellar core, a non-negligible effect at the masses and couplings we are considering here.
Consequently, our result modifies considerably the previous constraint, especially for $m_a \gtrsim 50$~keV.
The combined constraints from Globular Cluster stars, 
SN 1987A, and beam-dump experiments 
 leave  a small triangularly shaped region open in the parameter space around
 $m_a \sim 0.5-1\,$ MeV and 
 $g_{a\gamma} \sim 10^{-5}$ {GeV}$^{-1}$.
 This is 
informally known as the ALP ``cosmological triangle'' since it can be excluded only
using standard cosmological arguments. 
As we shall mention, however, there are viable cosmological models that are compatible with axion-like particles with 
parameters in such region.
We also discuss possibilities to explore the cosmological triangle experimentally in upcoming accelerator experiments.
\end{abstract}

\maketitle

\section{Introduction}

Axion-like-particles (ALPs) with masses $m_a$ in the  keV-MeV range  emerge in different extension of the Standard Model, 
as Pseudo-Goldstone bosons of some broken global symmetry. 
The theoretical speculation about \emph{superheavy} axion models began long ago
(see Sec.~6.7 of Ref.~\cite{DiLuzio:2020wdo} for a recent review), in an attempt to get rid of 
the strong astrophysical bounds on the axion coupling, which made it effectively invisible.
In this context, superheavy means heavier than about $100\,{\rm keV}$, so that the axion 
production in most stars (supernovae and neutron stars being an exception) 
is Boltzmann suppressed and the majority of the stellar axion bounds are
relaxed.
Nowadays, several mechanisms exist to increase the axion mass independently from 
its couplings, without spoiling the solution of the strong CP problem 
(a list of references can be found in~\cite{DiLuzio:2020wdo}).

Besides QCD axions, heavy ALPs emerge in compactification scenarios of  string theory~\cite{Svrcek:2006yi,Arvanitaki:2009fg,Cicoli:2012sz}, or in the context of  ``relaxion''
models~\cite{Graham:2015cka}.
Heavy ALPs have also recently received considerable attention in the context of Dark Matter model-building. 
Indeed, they may act as mediators for the interactions between
the  Dark Sector and Standard Model (SM) allowing to reproduce the correct Dark Matter relic abundance via thermal freeze-out~\cite{Boehm:2014hva,Hochberg:2018rjs}. 
ALPs with masses below the MeV scale can have a wide range of implications for cosmology and astrophysics (see~\cite{Dolan:2017osp}
for a review), affecting for example Big Bang Nucleosynthesis (BBN), 
the Cosmic Microwave Background (CMB)~\cite{Cadamuro:2010cz,Cadamuro:2011fd,Depta:2020wmr} and the evolution of stars. 
Colliders and beam-dump experiments are also capable to explore this mass range, 
indeed reaching the $m_a \sim \mathcal{O}$(GeV) frontier, which is not covered by 
any astrophysical or cosmological considerations~\cite{Dolan:2017osp,Jaeckel:2015jla,Dobrich:2019dxc}. 

In this work we are interested in ALPs interacting exclusively with photons.
Additional couplings with SM fields, particularly  with electrons, may spoil some of our conclusions. 
For such ALPs, the collection of all the astrophysical and experimental constraints 
leaves  a triangular area in the parameter space, for masses 
$m_a \sim 0.5-1\,$ MeV and 
 couplings $g_{a\gamma} \sim 10^{-5}$ {GeV}$^{-1}$, open.
Although the existence of ALPs with such parameters is in tension with standard cosmological arguments~\cite{Cadamuro:2010cz,Cadamuro:2011fd},
the region of such masses and couplings passes the current experimental tests and all the known astrophysical arguments, 
and is also permitted in viable non-standard cosmological scenarios~\cite{Depta:2020wmr}.
Because of that, this parameter area is sometimes dubbed as the ALP \emph{``cosmological triangle''}.
As we shall discuss in Sec.~\ref{sec:Experiments}, 
this region is now the target of several direct investigations, as more and more 
experiments are reaching the sensitivity to probe those masses and couplings,
and there is a chance that such area might be covered in the next decade or so.
Redefining the boundaries of the cosmological triangle is, therefore, particularly timely and relevant to guide the experimental investigations.

In this work we revisit the globular cluster bound on heavy ALPs, 
which defines the low-mass boundary of the cosmological  triangle.
Globular Clusters (GC) are gravitationally bound systems of stars, typically harboring  
a few millions stars.
Being among the oldest objects in the Milky Way, their population is made of low-mass stars ($M<1 M_\odot$). Most of these stars belong to the so-called Main Sequence, which corresponds to the  H burning evolutionary phase. However, there are two other well defined evolutionary phases, i.e., the Red Giant Branch (RGB) and the Horizontal Branch (HB).
The first is made by cool giant stars, burning H in a thin shell surrounding a compact He-rich core. During the RGB phase the stellar luminosity increases and the core contracts, until the temperature rises enough to ignite He. Then, stars leave  the RGB and enter the HB phase, during which they burn He, in the core, and H, in the shell. 

 The number of stars found  in the different evolutionary phases depends linearly on 
the time spent by a star in each of them. For this reason, stellar counts provide a powerful tool to investigate the efficiency of the energy sources and sinks in stellar interiors, those that affect the stellar lifetime $\tau$ in a given stage of the stellar evolution.
In this context, the GC $R$ parameter, 
defined as the number ratio of horizontal branch  to red giants branch stars, i.e.:
\begin{equation}
R= \frac{N_{\rm HB}}{N_{\rm RGB}}=\frac{\tau_{\rm HB}}{\tau_{\rm RGB}},
\label{erre}
\end{equation}
\noindent
is a powerful observable often used to investigate stellar physics. 
In particular, it has been also exploited to constrain the axion-photon coupling $g_{a\gamma}$~\cite{Raffelt:1987yu,Raffelt:2006cw,Ayala:2014pea,Straniero:2015nvc},
at least for ALPs light enough, $m_a\lesssim 30\,{\rm keV}$, that their production is not Boltzmann suppressed.
At such low masses, the most relevant axion production mechanism induced by the photon coupling 
is the Primakoff process, $\gamma + Ze \to Ze +a$, i.e. the conversion of a photon into an 
ALP in the electric field of nuclei and electrons in the stellar plasma (cf. Sec.~\ref{sec:emissivity}). 
This process is considerably more efficient in HB than in RGB stars,
since in the latter case it is suppressed due to the larger screening scale and plasma frequency (see Sec.~\ref{sec:emissivity}). 
Therefore, the energy-loss caused by the production of ALPs with a sizable  $g_{a\gamma}$ would imply a reduction of the HB lifetime and, in turn, a reduction of the $R$ parameter.
As it turns out, $R$ has a substantial dependence, approximately linear, on the helium abundance of the cluster and, 
if ALPs are also included, a quadratic dependence on the axion-photon coupling. 
On the other hand, the $R$ parameter is only marginally affected by a variation of the cluster age and metallicity. 
 Thus, once the He abundance is known from direct or indirect measurements, bounds (or hints) 
 on the axion-photon coupling can be obtained from the comparison of the $R$ parameter measured in 
 Globular clusters with the theoretical expectations obtained by varying $g_{a\gamma}$~\cite{Ayala:2014pea,Straniero:2015nvc,Giannotti:2015kwo,Giannotti:2017hny}. 
An accurate application of this method,  based on photometric data for 39 GCs, was discussed in~\cite{Ayala:2014pea} by some of us,  who found
an upper bound $g_{a\gamma} < 0.66  \times 10^{-10}$  {GeV}$^{-1}$ at 95 \% confidence level, a value more recently experimentally confirmed by the CAST
 collaboration~\cite{Anastassopoulos:2017ftl}. 

The goal of the present work is to extend the GC bound on $g_{a\gamma}$  to higher axion masses. 
Given the typical temperature  $T\sim 10\,$keV in the stellar core of a HB star, 
one expects the thermal production of particles to be Boltzmann suppressed for $m_a \gtrsim 30\,$keV, 
relaxing the bound on  $g_{a\gamma}$. 
A quantitative analysis was carried out in~\cite{Cadamuro:2011fd}, where a bound was derived 
from the requirement that the axion energy emitted per unit time and mass, $\varepsilon_a $, averaged over a typical HB core, satisfies the requirement 
$\langle \varepsilon_a \rangle \lesssim 10 \, \textrm{erg}\,\textrm{g}^{-1} \,\textrm{s}^{-1}$~\cite{Raffelt:1996wa}.
However, that analysis neglected the contribution of the photon coalescence process 
$\gamma\gamma \to a$ (cf. Sec.~\ref{sec:emissivity}),
to the ALP production in stars. 
At low masses, this process is subdominant and it is forbidden for $m_a<2\,\omega_{\rm pl}$, 
where $\omega_{\rm pl}$ is the plasma frequency at the position where the process takes place.
Hence, the inclusion of the photon coalescence does not affect the bound obtained in Ref.~\cite{Ayala:2014pea}, 
which remains valid for light axions ($m_a \lesssim 10$ keV). 
As we shall show in Sec.~\ref{sec:emissivity}, however, for masses $m_a \gtrsim 50\,{\rm keV}$, 
the coalescence production dominates and becomes several times larger than the Primakoff at masses $\gtrsim 100 \,{\rm keV}$.
Furthermore, ALPs with a large mass and coupling have a non-negligible probability to decay inside the stellar core.
In this case, they would not contribute to the cooling of the star. 
We show that this is the case for the couplings and masses within the cosmological triangle and conclude that the stellar bounds in this region are considerably relaxed with respect to what shown in the previous literature.

The plan of our work is the following. In Sec.~\ref{sec:emissivity}, we revise the axion emissivity via the
Primakoff conversion and the photon coalescence.
In Sec.~\ref{sec:GC_bound}, we discuss our procedure and present our bound on  $g_{a\gamma}$ for massive ALPs. 
Then, we  show the complementarity of our 
bound with other constraints.
In Sec.~\ref{sec:SN}, with that from SN 1987A (in the trapping regime),
and in Sec.~\ref{sec:Experiments} with the experimental bounds from beam-dump searches.
Finally, in Sec.~\ref{sec:conclu} we summarize our results and we conclude.
In Appendix~\ref{app:primakoff} we compute the photon-axion transition rate from Primakoff conversion and in \ref{app:coalescence} we calculate the ALP production rate from Primakoff conversion and photon coalescence.

\section{Axion emissivity}
\label{sec:emissivity}

The ALP-two photon vertex is described by the Lagrangian term
\begin{equation}
\label{mr}
{\cal L}_{a\gamma}=-\frac{1}{4} \,g_{a\gamma}
F_{\mu\nu}\tilde{F}^{\mu\nu}a=g_{a\gamma} \, {\bf E}\cdot{\bf B}\,a~,
\label{eq:lagrangian}
\end{equation}
where $g_{a\gamma}$ is the ALP-photon coupling constant (which has dimension of an inverse energy),
$F$ the electromagnetic field and $\tilde F$ its dual.

The primary production mechanisms for ALPs interacting with {\emph{transverse} photons in the core of a HB star are:
\begin{itemize}
\item the Primakoff conversion $\gamma + Ze \to Ze +a$, where a thermal photon in the stellar core 
converts into an axion in the Coulomb fields of nuclei and electrons; 
\item the photon coalescence process $\gamma \gamma \to a$, where two photons in 
a medium of sufficiently high density annihilate producing an axion.  
\end{itemize}
As we shall see, the former dominates at low ALP masses ($m_a \lesssim 50$~keV) while at large mass 
the photon coalescence takes over. 

There is a vast literature on the axion Primakoff conversion rate.
The interested reader may consult Ref.~\cite{Raffelt:1987yu,Cadamuro:2011fd}
for a detailed discussion. 
Here, we provide only a brief review and present some results applicable in the typical plasma conditions relevant for this work.
In general, the axion emission rate (energy per mass per time) via the Primakoff conversion is given by the expression
\begin{equation}
\varepsilon_a = \frac{2}{\rho} \int \frac{dk \,k^2}{2 \pi^2} \Gamma_{\gamma\to a}\, E\, f(E) \,\ ,
\end{equation}
where the factor 2 comes from the photon degrees of freedom,
$ \rho $ is the local density, 
 $f(E) = (e^{E/T}-1)^{-1}$ is the Bose-Einstein distribution, and 
$\Gamma_{\gamma \to a}$ is the photon-axion transition rate,
\begin{eqnarray}
\Gamma_{\gamma \to a} &=& \frac{g_{a\gamma}^2 T \kappa^2}{32 \pi}{\frac{p}{E}}
\bigg\{ \frac{[(k+p)^2 + \kappa^2][(k-p)^2 + \kappa^2]}{4 p k \kappa^2} \nonumber \\
& & \ln\bigg[\frac{(k+p)^2 + \kappa^2}{(k-p)^2 + \kappa^2} \bigg] \nonumber \\
&-&\frac{(k^2-p^2)^2}{4kp\kappa^2} \ln\bigg[\frac{(k+p)^2}{(k-p)^2} \bigg] -1\bigg\} \,\ .
\label{eq:primakoff}
\end{eqnarray}
In the last expression, $E$  and $p=\sqrt{E^2-m_a^2}$ are, respectively, the ALP energy and momentum.
The photon obeys the dispersion relation $k=\sqrt{\omega^2-\omega_{\rm pl}^2}$ where
$k$ is the photon momentum, $\omega$ its energy, and $\omega_{\rm{pl}}^2 \simeq 
4\pi \alpha n_e/m_e$ is the plasma frequency (or effective ``photon mass''). In a photon-axion transition the energy is conserved because we ignore recoil effects. Therefore, we use $\omega = E$.
Finally $\kappa$ is the screening scale 
 \begin{equation}
 \kappa^2 = \frac{4 \pi \alpha}{T}\left(n_e^{\rm eff}+ \sum_j Z_j^2 n_j^{\rm eff} \right) \,\ ,
 \label{eq:screen}
 \end{equation}
where $n_e^{\rm eff}$ and $n_j^{\rm eff}$ are, respectively, the effective number of electrons and ions with nuclear charge $Z_j\,e$. 
Note that in the centre of a HB star, $T \sim 8.6$ keV, 
$\rho \sim 10^4$ g cm$^{-3}$, and 
$\omega_{\rm pl} \sim 3$ keV. 
Thus, 
the plasma frequency is considerably smaller than the thermal energy. 
Nevertheless, to achieve a higher accuracy 
our numerical Primakoff emission rate includes also the effects induced by a finite plasma frequency 
(a detailed description of the adopted emission rate
can be found in the Appendix of Ref.~\cite{Straniero:2019dtm}, which we have generalized at finite ALP mass).

The axion coalescence process, $\gamma\gamma\to a$, 
has a kinematic threshold, vanishing for $m_a\leq 2\omega_{\rm pl}$~\cite{DiLella:2000dn}.
As we shall see, above this threshold the production rate is a steep function of the mass
and dominates over the Primakoff at  $m_a\gtrsim 50\,$ keV.
In order to calculate the axion coalescence rate in a thermal medium, it is convenient  to approximate the Bose-Einstein photon distribution with a Maxwell-Boltzmann, $f(E)\to e^{-E/T}$, for the photon occupation number~\cite{DiLella:2000dn}. 
This approximation is justified since we are interested only in axion masses 
(and thus axion energies) of the order of the temperature or larger 
(for $m_a\lesssim T$ the coalescence process is practically negligible). 
As shown in Appendix~\ref{app:coalescence}, the production rate per unit volume of ALPs of energy between $E$ and
$E+dE$ is~\cite{DiLella:2000dn}
\begin{equation}
d \dot N_a  = \frac{g_{a \gamma}^2 m_a^4}{128 \pi^3}p\left(1- \frac{4 \omega_{\rm pl}^2}{m_a^2} \right)^{3/2} e^{-E/T} dE \,\ ,
\end{equation}
and the axion emissivity (per unit mass):  
\begin{equation}
\varepsilon_a = \frac{1}{\rho}\int  E \frac{d \dot N_a}{d E} dE \,\ .
\end{equation}

\begin{figure}[t!]
\vspace{0.cm}
\includegraphics[width=1.\columnwidth]{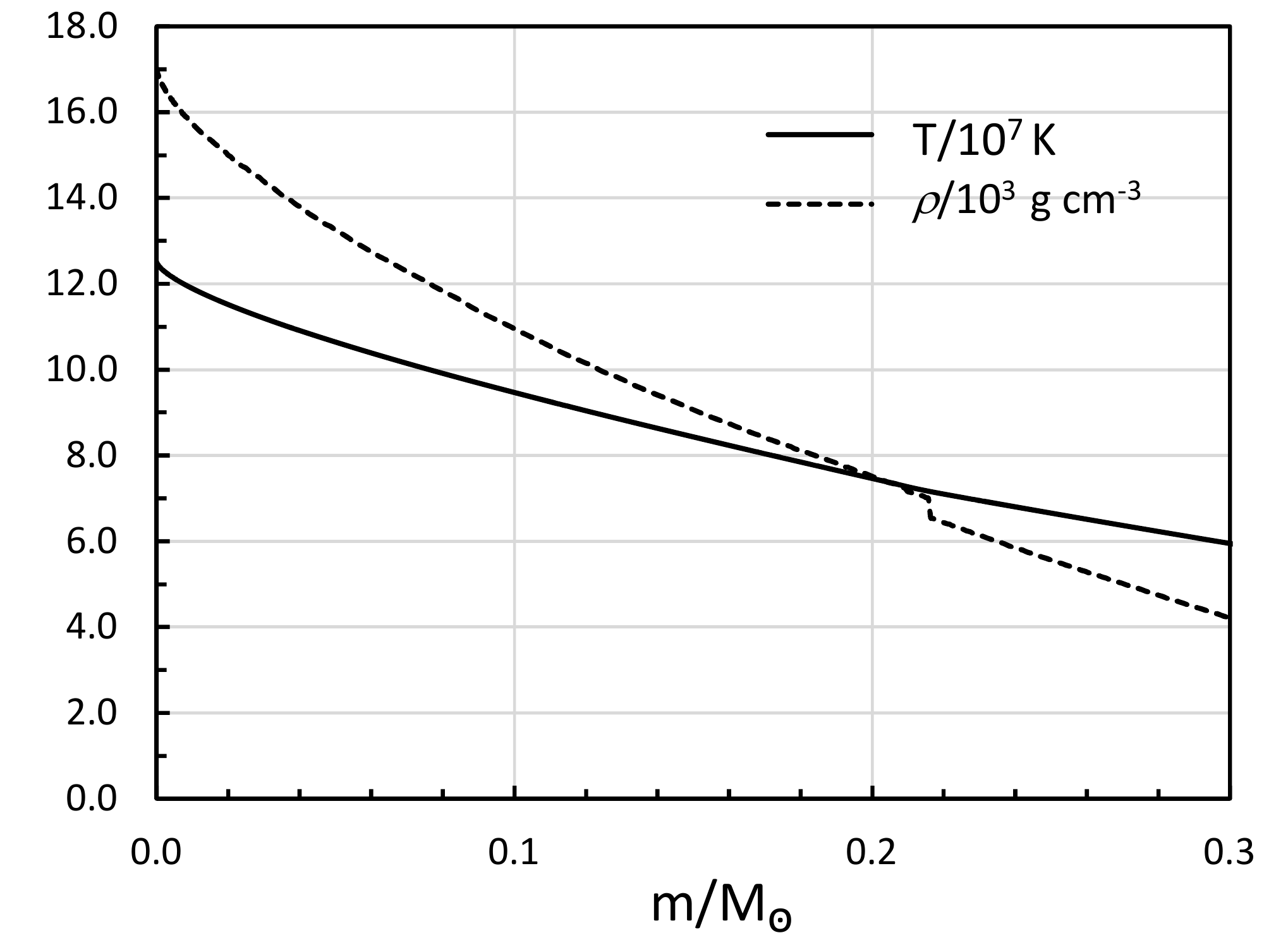}
\caption{Profiles of temperature $T$ (solid line) and density $\rho$ (dotted line)
within the core of typical HB star (see text for details). 
The most internal 0.3 $M_\odot$ are shown.
 }
\label{fig:HB_model}
\end{figure}

The temperature and density profiles within the He-rich core of a typical HB stellar model 
are shown in Fig.~\ref{fig:HB_model}.
The model has been evolved starting from the pre-main sequence up to the end of the core He burning phase. 
For the initial structure ($t=0$ model) we have adopted a mass M=0.82 $M_\odot$ and, as usual, 
a homogeneous composition, namely: $Y=0.25$ and $Z=0.001$. 
After $\sim 13$ Gyr the central He burning begins (zero age HB). At that time the stellar mass
is $m \sim 0.72$ $M_\odot$, while the mass of the He-rich core is $m \sim 0.5$ $M_\odot$. 
Fig.~\ref{fig:HB_model} is a snapshot of the stellar core taken when the central mass fraction of He reduces down to $X_{He}\sim 0.6$.  
The corresponding Primakoff and photon coalescence emission rates are compared in Fig.~\ref{fig:energyloss}.
The quantity reported in the vertical axis is the ratio of the energy-loss rate,
in units of erg$\,$g$ ^{-1} \,$s$ ^{-1} $, 
and the square of the axion-photon coupling, $g_{10}\equiv g_{a\gamma}/10^{-10}$  {GeV}$^{-1}$. The Primakoff and photon coalescence emission rates have been computed 
for two different values of the axion mass, namely:   $m_a=30$~keV  and $m_a=80$~keV. 
In the case of $m_a=30$~keV, the Primakoff energy-loss rate (in the center of the star) is a factor of $\sim 3$ larger than the photon coalescence rate. 
Conversely, for  $m_a=80$~keV the photon coalescence dominates and the contribution of the Primakoff is effectively negligible.

Fig.~\ref{fig:lumen} shows how the ALP luminosity,
\begin{equation}
L_a = 4 \pi \int  \rho \varepsilon_a r^2 dr \,\ ,
\end{equation}
\noindent
 depends on the ALP mass $m_a$.  The integration is extended from the center ($r=0$) to the stellar surface. 
According to our expectations, the coalescence process is sub-leading for $m_a \lesssim 50$~keV, but it dominates at higher masses.
Note that the stellar luminosity, as due to the photon emission,
 is $L\sim 2\times 10^{35}$ erg s$^{-1}$, which is about 2 orders of magnitude larger than the 
axion luminosity at $g_{10}\sim 1$.

\begin{figure}[t!]
\vspace{0.cm}
\includegraphics[width=1.\columnwidth]{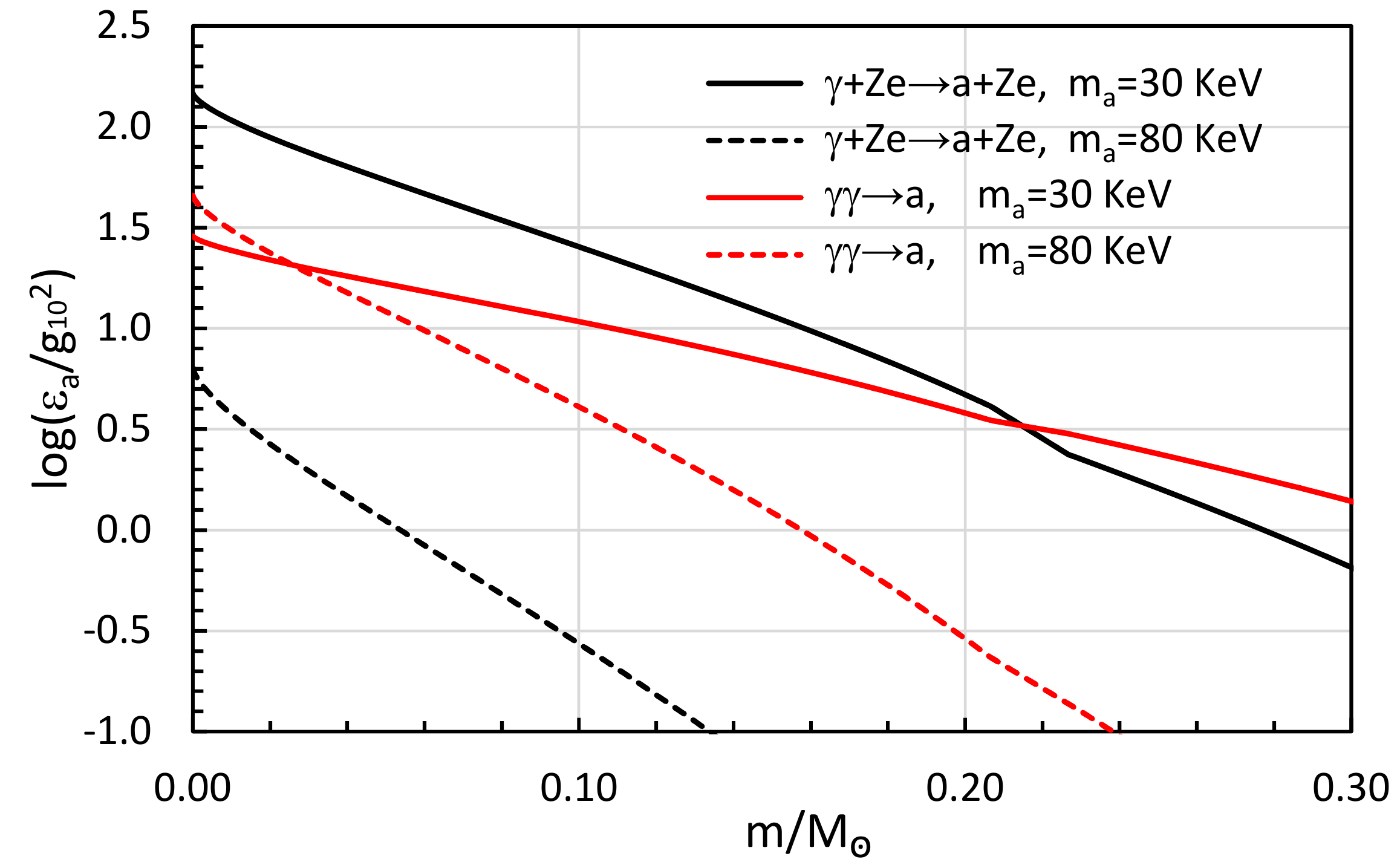}
\caption{Energy-loss rates (in units of erg$\,$g$ ^{-1} \,$s$ ^{-1} $ and normalized for $g_{10}= 1$)	
for Primakoff  ($\gamma+Ze\to a+Ze$) and photon coalescence 
($\gamma\gamma\to a$) within the core of a typical HB star. The most
internal 0.3 $M_\odot$ is shown. This is the same model used for Fig \ref {fig:HB_model}.
Results for two different axion mass, $m_a=30$~keV and $m_a=80$~keV, are shown.
}
\label{fig:energyloss}
\end{figure}

\begin{figure}[t!]
\vspace{0.cm}
\includegraphics[width=1.\columnwidth]{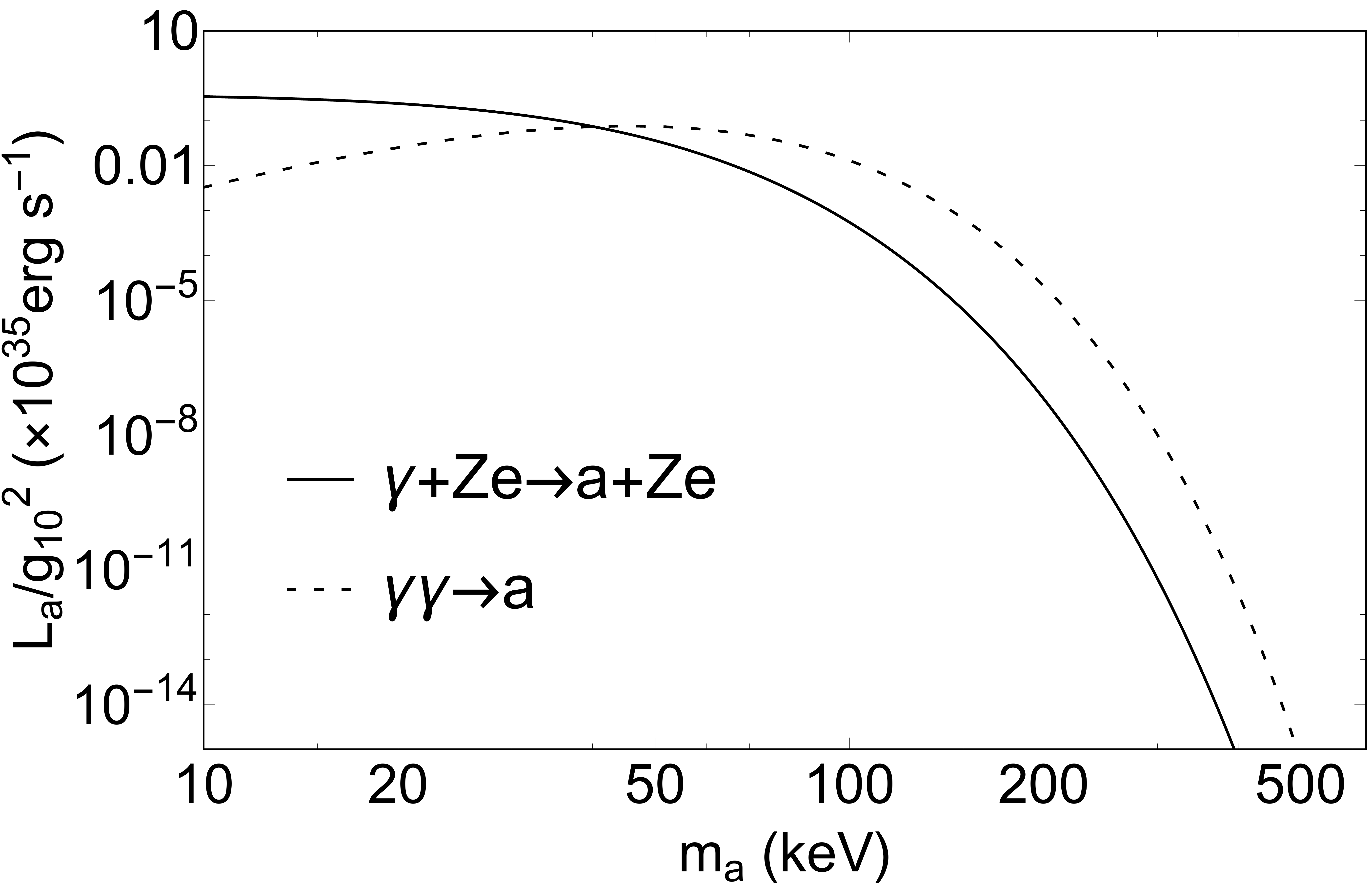}
\caption{ALP luminosity for Primakoff process ($\gamma+Ze\to a+Ze$, continuous curve) and for 
photon coalescence ($\gamma\gamma\to a$, 
dashed curve) versus axion mass $m_a$. The HB model is the one used in 
Fig. \ref{fig:HB_model}. As for the rates in Fig. \ref {fig:energyloss}, 
the luminosities are normalized to $g_{10}= 1$. }
\label{fig:lumen}
\end{figure}

\section{Globular Cluster bound}
\label{sec:GC_bound}

In order to derive a bound on $g_{a \gamma}$ for massive ALPs, we have computed several
evolutionary sequences of stellar models, from the pre-main-sequence to the end of the core He burning. The models have been computed by means of FuNS (Full Network Stellar evolution), an hydrostatic 1D stellar evolution
code~\cite{Straniero:2019dtm}.
In general, the inclusion of the axion energy-loss in stellar model computations leads to a reduction of the 
$R$ parameter, defined in Eq.~(\ref{erre}). 
On the other hand, the larger the initial He abundance the larger the estimated $R$. 
In practice, the upper bound  on the axion-photon coupling is obtained when the largest possible value of the He abundance is assumed. 
Analyzing the He abundance measured in molecular clouds with metallicity in the same range of those of galactic GCs, 
in Ref.~\cite{Ayala:2014pea} it was estimated a conservative upper limit for the He abundance, specifically $Y=0.26$. 
Adopting this value of $ Y $, it was shown that the $ R $ parameter obtained from photometric observations of 39 GCs,
$ R = 1.39\pm0.03 $,  implies the stringent  upper bound $ g_{a\gamma}= 0.66 \times 10^{-10}\, {\rm GeV^{-1}} $ (95 \% C.L.). 
However, this bound is only valid for light axions.

 
 Since ALPs interacting only with photons are not efficiently produced  in the core of RGB stars, 
  and hence affect minimally the RGB lifetime ($\tau_{RGB}$ in Eq.~(\ref{erre})), 
  the variation of $R$ due to an axion production is essentially a consequence of the reduction of the HB lifetime ($\tau_{HB}$ in Eq.~(\ref{erre})).  
We have computed $\tau_{HB}$  for a GC benchmark.
Specifically, we used: age $13$ Gyr, metallicity $Z=0.001$, and $Y=0.26$, corresponding to the conservative upper limit for the GC He abundance reported in~\cite{Ayala:2014pea}. 
In the standard case, when no exotic energy-loss process is included, we found $ \tau_{HB}= 8.84 \times 10^{7} \,\ \textrm{yr}$.
The addition of light axions with $g_{a\gamma} = 0.66  \times 10^{-10}$  {GeV}$^{-1}$ reduces the HB lifetime down to 
 $ \tau_{HB}= 7.69 \times 10^{7} \,\ \textrm{yr}$.
Requiring the HB lifetime to be within these values guarantees that 
the predicted $R$ parameter is consistent, within $ 2\,\sigma $, with the observed one. 

The argument was generalized to massive ALPs by searching for the ALP-photon coupling 
that reduces the HB lifetime down to $ \tau_{HB}= 7.69 \times 10^{7} \,\ \textrm{yr}$ at each fixed ALP mass.  
Notice that the ALP decay length decreases rapidly with the ALP mass and coupling,
\begin{equation}
\lambda_a= 5.7 \times 10^{-5} \,\ g_{10}^{-2}m_{\rm keV}^{-3}\frac{\omega}{m_a}\sqrt{1-\left(\frac{m_a}{\omega}\right)^{2}} R_{\odot} \,\ ,
\end{equation}
with $m_{\rm keV}=m_a/(1 \,\ \textrm{keV})$. 
Thus, for the masses and couplings we are interested in, 
a considerable fraction of ALPs may decay inside the star.
 Those ALPs do not contribute to energy loss, but they can lead to an efficient
 energy transfer inside the star~\cite{Raffelt:1988rx}. 
 In order to address this issue one should perform a dedicated simulation of HB evolution including
 ALP energy transfer. 
 This is a challenging task that we leave for a future work. 
 For the moment we adopt a conservative approach assuming that
 the ALPs decaying inside the \emph{convective} core, with a radius $R_c\simeq 3\times 10^{-2} R_{\odot}$,  
 do not lead to any energy transfer, convection being a very efficient energy transfer mechanism by itself.
 Neglecting the contribution of these ALPs leads to the deterioration of the ALP bound that we observe for
 $g_{a\gamma} \gtrsim 10^{-6}$ GeV$^{-1}$. 
 Our result remains effectively unchanged if we replace the convective core with the entire Helium core,
$R\simeq 7\times 10^{-2} R_{\odot}$, as our threshold radius.
We stress, however,  that our bound might relax even
further  if a detailed simulation were to show that even ALPs decaying at larger radii are inefficient in transferring energy.
Our result is shown in the exclusion plot reported in Fig.~\ref{fig:bound}.
The continuous red line indicates our new result (95 \% C.L.) 
while the dashed gray line represents the bound ignoring the coalescence production and the ALP decay, 
and corresponds roughly to the previous constraint.
It is evident how the bound loses its strength 
for masses above  $\sim 30$ keV, because of the Boltzmann suppression of the axion emissivity. 

For such high masses one may ask if ALPs can be gravitationally trapped into the star gravitational field. In this case ALPs escape only if their kinetic energy is greater than
\begin{equation}
U(r)=\frac{G M_{r}m_{a}}{r}=7.44\times10^{-34}{\rm keV}\frac{M_{r}}{{\rm g}}\frac{m_{a}}{{\rm keV}}\frac{{\rm km}}{r}\;;
\end{equation}
where $M_{r}$ is the star mass up to the radius $r$ and $m_{a}$ is the ALP mass. As a simple estimate we consider the border of the core, outside the gravitational potential well is weaker. Therefore we use $M_{r}=10^{33}~{\rm g}$, $r=5\times10^{4}~{\rm km}$ and $m_{a}=500~{\rm keV}$ obtaining $U(r)=8\times10^{-3}~{\rm keV}$ which is much smaller than the typical temperature $T\sim10~{\rm keV}$. In conclusion this effect is negligible.

For reference, in the figure we are also showing, in light green,
the region excluded by SN 1987A in the regime of ALPs trapped in the SN core (see Sec.~\ref{sec:SN}),
 and in blue the parameters excluded by direct searches at beam dump experiments (see Sec.~\ref{sec:Experiments}).
 %
\begin{figure}[t!]
\vspace{0.cm}
\includegraphics[width=1\columnwidth]{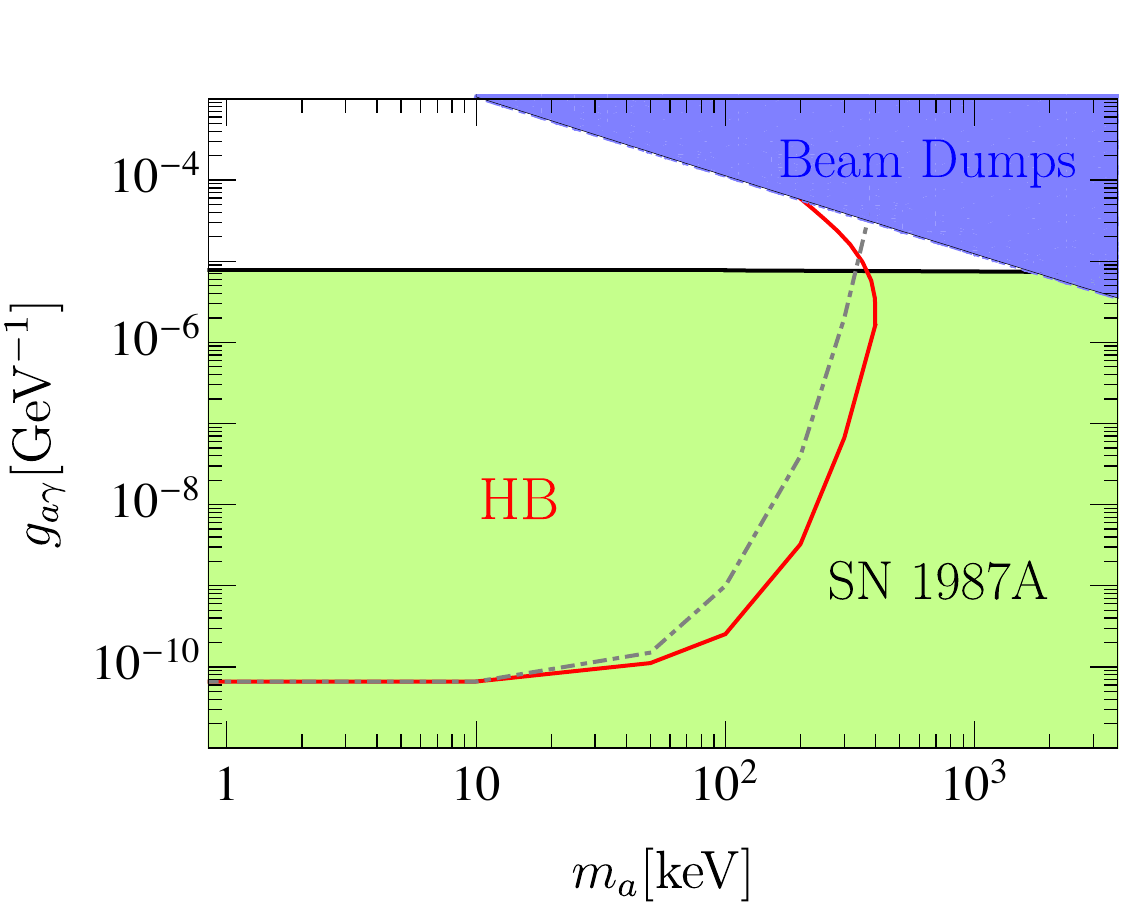}
\caption{HB bound (red line) in the plane $g_{a\gamma}$ \emph{vs} $m_a$, compared with other exclusion limits.
The dashed gray curve presents the HB limit accounting only for Primakoff while the continuous red curve includes
also the photon coalescence process.}
\label{fig:bound}
\end{figure}
%
Interestingly, the combination of all the astrophysical and experimental bounds 
 leave a small triangular area, roughly at $m_a \sim 0.5-1\,$MeV and 
 $g_{a\gamma} \sim 10^{-5}$ {GeV}$^{-1}$, unconstrained. 
 This is the ALP cosmological triangle.
 Standard cosmological arguments, particularly concerning BBN and the allowed
 effective number of relativistic species, $N_{\rm eff}$, 
 can be used to exclude this area~\cite{Cadamuro:2011fd,Depta:2020wmr}. 
 Nevertheless, in non-standard cosmological scenarios, e.g. in low-reheating models,
 the cosmological bounds can be relaxed all the way to the GC bound calculated in this work~\cite{Depta:2020wmr}.
 Thus, the cosmological triangle is still a viable region of the ALP parameter space, open to  
 experimental and phenomenological investigations. 

\section{SN 1987A  bound from axion trapping} 
\label{sec:SN}

For the sake of completeness, in this section we present briefly 
our derivation of the SN 1987A constraint on heavy ALPs presented in Fig.~\ref{fig:bound} and \ref{fig:overview}.
A detailed study of this constraint, based on state-of-the-art SN models~\cite{Fischer:2016boc},
is currently ongoing and will be the topic of a forthcoming work by some of us~\cite{lucente}.
Here, we just present a succinct discussion of the SN argument to constraint the ALP-photon couplings at the bottom edge of the 
cosmological triangle. 
In order to characterize the ALP emissivity in a SN, and in particular the effect of degeneracy in a SN core 
we closely follow~\cite{Payez:2014xsa}.

Heavy ALPs can be copiously produced in a supernova (SN) core via Primakoff and coalescence processes.
Due to the higher core temperature, $T\sim {\mathcal O}(30)$~MeV, 
SNe can be used to probe ALP masses considerably larger than those probed by GCs (see, e.g.~\cite{Dolan:2017osp,Lee:2018lcj,Ertas:2020xcc}). 
For couplings of interest in this work, $g_{a\gamma} \sim {\mathcal O}(10^{-5})$~GeV$^{-1}$, 
ALPs would be trapped in the SN, 
having a mean-free path smaller than the size
of the SN core ($R \sim 10$~km)~\cite{Dolan:2017osp,Lee:2018lcj}.
In this case, ALPs may contribute significantly to the energy transport in the star, modifying the SN evolution.
Since SN 1987A neutrino data are in a reasonable agreement with core-collapse SN models without the emission of exotic species, 
one should require that ALPs  interact more strongly than the particles which provide the standard mode of energy transfer, i.e. neutrinos.

When ALPs interact strongly enough to be trapped in the SN core, they are
emitted from an \emph{axion-sphere}, a spherical shell whose radius $r_a$ is fixed by the optical depth being about unity.
More specifically, we calculated $ r_a $ imposing that the optical
depth
\begin{equation}
\tau_a = \int_{r_a}^{+\infty} \kappa_a \rho dr \,\ ,
\label{eq:tau}
\end{equation}
where $\kappa_a$ is the axion opacity, 
satisfies the condition $\tau_a(r_a) \simeq 2/3$. 
This is analogous to the neutrino last scattering surface, i.e. the  ``neutrino-sphere'', with radius $r_\nu$.

Trapped ALPs have a black-body emission with a luminosity
$L_a \propto r_a^2 T^4(r_a)$. 
In order to obtain the bound on $g_{a\gamma}$ one should impose~\cite{Raffelt:1996wa,Raffelt:1987yt}
\begin{equation}
L_a \lesssim L_\nu \,\ .
\label{eq:luminanu}
\end{equation}

We are concerned mostly with a time posterior to 
0.5--1~s, where the outer core has settled
and the shock has begun to escape.
Specifically, in our numerical calculation we refer to the SN model used in~\cite{Carenza:2019pxu}, for a representative post-bounce time 
$t_{\rm pb}=1$~s.


We calculated the ALP opacity following the prescriptions in~\cite{Raffelt:1988rx}
(see \cite{Ertas:2020xcc} for an alternative approach).
For masses $m_a \lesssim$ a few MeV, the dominant contribution  to the axion
opacity is due to the inverse Primakoff conversion, $a +Ze \to \gamma +Ze $, 
\begin{equation}
\kappa_{a\to \gamma} = \frac{1}{\rho \lambda_{a\to \gamma}} = \frac{1}{\rho} \frac{\Gamma_{a\to \gamma} }{\beta_E} \,\ ,
\end{equation}
where $\lambda_{a\to \gamma}$ is the mean free-path, 
and $\beta_E = (1-m_a^2/E^2)^{1/2}$.
The inverse Primakoff conversion rate is 
$\Gamma_{a\to \gamma }= 2 \Gamma_{\gamma \to a}$, with $\Gamma_{\gamma \to a}$ given in Eq. (\ref{eq:primakoff}).

From $\kappa_{a\to \gamma}$ one can calculate the mean ALP Rosseland opacity~\cite{Raffelt:1996wa}
\begin{equation}
	\kappa_a^{-1}= \frac{\int_{m_a}^\infty \kappa_{a\to \gamma}^{-1} \beta_E \partial_T B_E dE}
	{\int_{m_a}^{\infty} \beta_E \partial_T B_E dE} \,\ ,
\end{equation}
where 
\begin{equation}
B_E = \frac{1}{2 \pi^2}\frac{E^2(E^2-m_a^2)^{1/2}}{e^{E/T} -1} \,\ , 
\end{equation}
is the ALP thermal spectrum.

We derived our bound on axion coupling from
the luminosity condition in Eq.~(\ref{eq:luminanu}),
taking the axion-sphere radius that satisfies Eq.~(\ref{eq:tau}).
As shown in Fig.~\ref{fig:bound}, for $m_a <10$~MeV, the luminosity condition excludes the values of the photon-axion coupling
$g_{a\gamma} \lesssim 8 \times 10^{-6}$~GeV$^{-1}$, in agreement with previous results~\cite{Dolan:2017osp,Lee:2018lcj}.

Note that the SN 1987A bound should not be considered at the same level of confidence as the GCs one, since it is not based on a self-consistent SN simulations.
Performing such a simulation, which should include also the trapped ALPs, 
would be a challenging task (see, e.g., \cite{DeRocco:2019jti}
for a recent investigation in the context of dark photons), and demand a separated investigation.

\section{Direct experimental tests of the cosmological triangle}
\label{sec:Experiments}

\begin{figure*}[t!]
	\vspace{0.cm}
	\includegraphics[width=1.8\columnwidth]{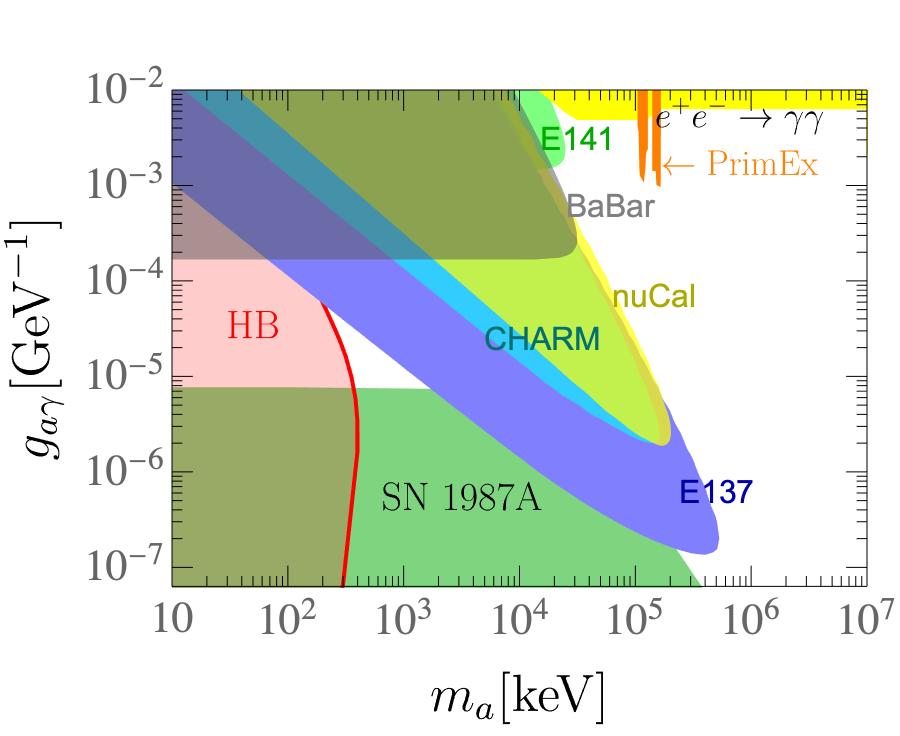}
	\caption{
		Overview of the heavy ALP parameter space in the plane $g_{a\gamma}$ \emph{vs} $m_a$. 
		The red-filled region labelled ``HB" represents our new exclusion result. 
		The SN 1987A bound~\cite{lucente} and the  experimental limits, compiled from Refs.~\cite{Dolan:2017osp,Dobrich:2019dxc}, are also shown.
		Prospects to experimentally probe the viable region are commented on in the text.
	}
	\label{fig:overview}
\end{figure*}

As discussed above, the ALP region at masses of a few MeV is the target of numerous investigations. 
In this section, we briefly comment on the existing experimental limits
near the cosmological triangle 
and future prospects to
test that region directly in experiments.

Fig.~\ref{fig:overview} shows an overview plot of the status of the 
search for heavy ALPs with our updated bound as discussed in this work in red, labelled ``HB''. Colored regions are excluded at 95\% C.L.
Other limits are compiled from references~\cite{Dolan:2017osp,Dobrich:2019dxc}
and detailed therein.
The experimental limits  which are 
 ``nose-like-shaped'' (E137~\cite{Bjorken:1988as}, CHARM \cite{Bergsma:1985qz}, nuCal  \cite{Blumlein:2011mv,Blumlein:2013cua}, E141 ~\cite{E141orig,Dobrich:2017gcm})  are from 
 beam-dump setups, in which the ALP needs to live long enough to reach
 the detection volume (boundary at ``large'' couplings and masses).
 However, it should not be so long-lived that it can excape from it 
 (boundary at ``small'' couplings and masses).
 
 The most efficient experiment to ``touch'' the cosmic triangle was E137,
 shown as a blue-shaded region in Fig.~\ref{fig:bound}.
 This bound is based on data
published by the experiment E137~\cite{Bjorken:1988as} and its revisit in~\cite{Dolan:2017osp}:
around 2$\times 10^{20}$ electrons were dumped into an aluminum target,
potentially
yielding to Primakoff-production of ALPs.
However, no excess of expected photon signals
was observed at a distance of $\sim 200$~m,
leading to an exclusion limit.

The small-coupling-limit of E137 relevant for us in this context
is largely determined by how {\it long-lived} 
ALPs can be while still being detected by the experiment.
The limit estimated~\cite{Dolan:2017osp} for this
reason seems robust
as late ALP decays will suffer little from their
non-negligible probability of showering
in air. We thus show this limit 
in Fig.~\ref{fig:overview}.
Roughly spoken, a  long baseline together
with a relatively soft ALP spectrum (compared to proton dumps
whose lower limits are at much larger couplings~\cite{Dobrich:2019dxc}),
made E137 an ideal fixed-target in probing the cosmological triangle at the top
section of its parameter space.

As for the possibilities to probe the remaining region at $m_a\sim 1\,$MeV,
Ref.~\cite{Dolan:2017osp} details on prospects to significantly
probe the cosmic triangle at Belle-II at a statistics
of 50~$\rm ab^{-1}$. Sensitivity is also expected at
``active'' beam dumps such as LDMX-type set-ups,
that can infer the presence of ALPs
through a ``missing-momentum signature''~\cite{Berlin:2018bsc}.
The running experiment PADME, at Frascati,
does not currently have the potential to reach the cosmological triangle~\cite{raggi,Alves:2017avw}
but could be potentially sensitive to this area after
a luminosity upgrade.

It is worth stressing that far
more experimental options to probe this triangle
exist if the axion-coupling is
not limited strictly to {\it direct} photon couplings~\cite{Ertas:2020xcc}.
However, this possibility is outside the assumptions 
made in our work.

\section{Discussions and conclusions}
\label{sec:conclu} 

In this work we have extended the GCs bound on the ALP-photon coupling to masses $m_a \gtrsim 10$~keV, in the region of the parameter space where the Boltzmann suppression of the axion emission rate can no longer be neglected. 
Our analysis improves on the previous work by including the coalescence process, $\gamma +\gamma\to a$, which is the dominating  axion production mechanism at masses above $\sim 50\,{\rm keV}$, 
and by accounting for ALPs decaying inside the stellar core. 
The bound is shown in Fig.~\ref{fig:bound} (red line), where we also compare it to the bound obtained ignoring the ALPs decay and the coalescence process (dashed gray line). 
The inclusion of the coalescence reduced the allowed value of the axion photon coupling by a factor of $\sim$ 4 at masses $\sim 100\,{\rm keV}$, and by over an order of magnitude at $ m_a\gtrsim 200\,{\rm keV} $.
At large masses and couplings, the ALP energy loss mechanism is hampered by ALPs decaying inside the stellar core and the axion bounds starts to relax. 
Quite interestingly, this effect becomes important very close to the edge of the cosmological triangle, 
opening up the region to future experimental probes.

Though excluded by
standard cosmological arguments,  
the cosmological triangle is a viable region in non-standard cosmological scenarios, e.g. in low-reheating models,
which relax substantially the cosmological bounds~\cite{Depta:2020wmr}.
Thus, it remains 
an area of great experimental interest, as shown in our Fig.~\ref{fig:overview}.
Indeed, several theoretical models permit ALPs (and even QCD axions) 
with parameters in this region, as discussed in Sec.~I,
making this a possible target area for future experimental investigations. 
Interestingly, a detection of an axion signal
 in this region would have dramatic cosmological consequences, 
 requiring non-standard cosmological scenarios.
 This intriguing possibility confirms once more the nice complementarity between astrophysical, cosmological arguments 
 and direct searches in order to corner or luckily discover axion-like-particles.

\section*{Acknowlegments}

We would like to thank Felix Kahlhoefer for helpful discussions.
We are also grateful to the anonymous referee for important comments concerning the role of ALP decays inside the star.
For this work, O.S has been funded  by the Italian Space Agency (ASI) and the Italian National Institute of Astrophysics (INAF) under the agreement n. 2017-14-H.0 -attivit\`a di studio per la comunit\`a scientifica di Astrofisica delle AlteEnergie e Fisica Astroparticellare.
The work of P.C. and 
A.M. is partially supported by the Italian Istituto Nazionale di Fisica Nucleare (INFN) through the ``Theoretical Astroparticle Physics'' project
and by the research grant number 2017W4HA7S
``NAT-NET: Neutrino and Astroparticle Theory Network'' under the program
PRIN 2017 funded by the Italian Ministero dell'Universit\`a e della
Ricerca (MUR).
B.D. acknowledges support through the European Research Council (ERC) under grant ERC-2018-StG-802836 (AxScale).
 \appendix

\section{Photon-axion transition rate from Primakoff conversion}
\label{app:primakoff}

	The differential rate for the Primakoff conversion is
	\begin{equation}
	d\Gamma_{\gamma\rightarrow a}=|\overline{\mathcal{M}}|^{2}\frac{V}{T}\frac{d^{3}\bf{p}}{(2\pi)^{3}}\;,
	\end{equation}
	where $V$ is the normalization volume, $T$ the interaction time and $|\overline{\mathcal{M}}|^2$ is the squared matrix element averaged over the initial photon polarization,
	\begin{equation}
	|\overline{\mathcal{M}}|^{2}=\frac{1}{2}|\mathcal{M}|^2=\frac{1}{2}\left|<a|\int dt d^3\textbf{r}\, g_{a\gamma}\,\phi_a\,\textbf{E}_e\cdot\textbf{B}|\gamma>\right|^{2}\,,
	\label{m}
	\end{equation}
	where $\phi_a$, $\textbf{B}$ and $\textbf{E}_e=\frac{Z\,e\,\textbf{r}}{|\textbf{r}|^3}$ are the interacting fields. By expanding the axion field $\phi_a$ and the magnetic field $\textbf{B}$ in plane waves, one obtains
	\begin{equation}
	|\overline{\mathcal{M}}|^2=\frac{1}{2}\left(\frac{g_{a\gamma}Ze}{2V}\right)^2\frac{\left|\textbf{k}\times\textbf{p}\right|^2}{|\textbf{k}-\textbf{p}|^4}\frac{2\pi T \delta(\omega_k-\omega_p)}{\omega_k\omega_p}\,.
	\label{mmedia}
	\end{equation}
	Therefore the transition rate results to be
	\begin{equation}
	\Gamma_{a\rightarrow \gamma}= \frac{1}{2V}\left(\frac{g_{a\gamma}Ze}{4\pi}\right)^2\frac{\left|\textbf{k}\times\textbf{p}\right|^2}{|\textbf{k}-\textbf{p}|^4}\frac{|\textbf{p}|}{E}\,d\Omega_p\,,
	\label{rate}
	\end{equation}
	where $E=\omega_k=\omega_p$ because of the delta function in Eq.~(\ref{mmedia}) and $\Omega_p$ is the scattering angle.\newline
	One has to consider that in a real plasma the particles mutually interact through their Coulomb fields and their motion is slightly correlated. This correlation
	implies the substitution \cite{Raffelt:1985nk}
	\begin{equation}
	\frac{1}{|{\bf{k}}-{\bf{p}}|^{4}}\rightarrow \frac{1}{|{\bf{k}}-{\bf{p}}|^{4}}\frac{|{\bf{k}}-{\bf{p}}|^{2}}{\kappa^{2}+|{\bf{k}}-{\bf{p}}|^{2}}\;,
	\end{equation}
	where  $\kappa$ is the screening scale in Eq.~(\ref{eq:screen}). 
	Thus one obtains 
	\begin{equation}
	\Gamma_{\gamma\rightarrow a}=g_{a\gamma}^{2}\frac{T\kappa^2}{32\pi^2} \frac{|\textbf{p}|}{E}\int d\Omega_p \frac{\left|\textbf{k}\times\textbf{p}\right|^2}{|\textbf{k}-\textbf{p}|^2(\kappa^2+|\textbf{k}-\textbf{p}|^2)}\,,
	\end{equation}
	and after an integration over the scattering angle we obtain Eq.~(\ref{eq:primakoff}).

\section{ALP production rate from photon coalescence}
\label{app:coalescence}
In order to obtain the ALP production rate from photon coalescence, let us consider the Boltzmann equation for the ALP distribution function $f_a$
\begin{eqnarray}
\frac{\partial f_{a}}{\partial t}=& &\frac{1}{2E}\int \frac{d^{3}\bf{k}_{1}}{(2\pi)^{3} 2\omega_{1}}\frac{d^{3}\bf{k}_{2}}{(2\pi)^{3} 2\omega_{2}} \nonumber \\
& &(2\pi)^{4}\delta^{4}(P-K_{1}-K_{2}) \frac{1}{2}|\overline{\mathcal{M}}|^{2} \nonumber \\
& &\left[(f_{a}+1)f_{\gamma} f_{\gamma}-f_{a}(f_{\gamma}+1)(f_{\gamma}+1)\right]\,,
\label{eq:boltzeq}
\end{eqnarray}
where $P=(E,\bf{p})$ is the ALP 4-momentum, $K_i=(\omega_i , \bf{k}_i)$ for $i=1,2$ are the 4-momenta of the two photons, and $|\overline{\mathcal{M}}|^2$ is the polarization-summed squared matrix element
\begin{equation}
|\overline{\mathcal{M}}|^{2}=\frac{1}{2}g_{a\gamma}^{2}m_{a}^{2}\left[m_{a}^{2}-4m_{\gamma}^{2}\right]\,.
\label{eq:transmatrix}
\end{equation}
The first term in Eq.~(\ref{eq:boltzeq}) describes the photon coalescence, while the second one is the decay process. Since one can assume that ALPs, once produced by photon coalescence, escape immediately, then $f_{a}=0$ and attention can be focused on the photon coalescence term
\begin{eqnarray}
\frac{\partial f_{a}}{\partial t}=& &\frac{1}{2E}\int \frac{d^{3}\bf{k}_{1}}{(2\pi)^{3} 2\omega_{1}}\frac{d^{3}\bf{k}_{2}}{(2\pi)^{3} 2\omega_{2}} (2\pi)^{4} \nonumber\\
&&\delta^{4}(P-K_{1}-K_{2})\frac{1}{2}|\overline{\mathcal{M}}|^{2}f_{\gamma}(\omega_{1}) f_{\gamma}(\omega_{2})\, .
\label{inverseterm}
\end{eqnarray}
By integrating, one obtains
\begin{equation}
\frac{\partial f_{a}}{\partial t}=
\frac{g_{a\gamma}^2m_{a}}{64\pi E_{a}} \left[m_{a}^2-4m_{\gamma}^2\right]^{3/2}e^{-E/T}\,,
\end{equation}
where a Maxwell-Boltzmann distribution for photons is assumed and $\omega_{1}+\omega_{2}=E$ because of the delta-function. Since
\begin{equation}
dN_a=f_a\frac{d^{3}\bf{p}}{(2\pi)^{3}}=\frac{f_{a}\, p \,E\, dE\, d\Omega}{(2\pi)^{3}}\,,
\end{equation}
the production rate per unit volume of ALPs of energy between $E$ and $E+dE$ results to be
\begin{equation}
\frac{d^{2} N_{a}}{dE\, dt}=\frac{g_{a\gamma}^{2}}{128\pi^{3}}m_{a}^4 \,p \left(1-\frac{4m_{\gamma}^{2}}{m_{a}^{2}}\right)^{3/2}e^{-E/T}\,.
\label{eq:dninverse}
\end{equation}

\end{document}